\documentclass[twocolumn,superscriptaddress,nobibnotes,nofootinbib,floatfix,aps,prl,citeautoscript,reprint]{revtex4}
\usepackage[pdftex]{color,graphicx}

\DeclareGraphicsExtensions{.jpg, .png , .pdf}
\usepackage{graphicx}
\usepackage{epsfig}
\usepackage{color}
\usepackage{latexsym}
\usepackage{amsmath,amssymb,bm}
\usepackage{natbib}
\usepackage{ulem}

\newcommand{\MPI}{Max-Planck-Institut f\"ur chemische Physik fester Stoffe, Dresden, Germany}
\newcommand{\ITA}{Institut f\"ur Theoret. und Angwandte Physik, Universit\"at Stuttgart, D-70550 Stuttgart, Germany}
\newcommand{\LPMCN}{LPMCN, UCBL, CNRS, UMR-5586, F-69622 Villeurbanne, Lyon, France}
\newcommand{\SIMAP}{SIMAP, UJF, CNRS, INP Grenoble, F-38402 St Martin d'H\`eres, France}
\newcommand{\LLB}{LLB, CEA, CNRS, UMR-12, CE-Saclay, F-91191 Gif-sur-Yvette, France}
\newcommand{\ISSP}{Institute of Solid State Physics, Vienna University of Technology, 1040 Vienna, Austria}
\newcommand{\UFR}{Physikalisches Institut,Johann Wolfgang Goethe-Universit\"at, D-60438 Frankfurt, Germany}

\begin{document}

\title{Phononic filter effect of rattling phonons in the thermoelectric clathrate Ba$_8$Ge$_{40+x}$Ni$_{6-x}$}

\author{H.\, Euchner}
\affiliation{\ITA}
\author{S. Pailh\`es}
\affiliation{\LPMCN}
\affiliation{\LLB}
\author{L. T. K. Nguyen}
\affiliation{\UFR}
\affiliation{\MPI}
\affiliation{\ISSP}
\author{W. Assmus}
\affiliation{\UFR}
\author{F. Ritter}
\affiliation{\UFR}
\author{A. Haghighirad}
\affiliation{\UFR}
\author{Y. Grin}
\affiliation{\MPI}
\author{S. Paschen}
\affiliation{\ISSP}
\author{M. de Boissieu}
\affiliation{\SIMAP}
\date{\today}

\begin{abstract}
One of the key requirements for good thermoelectric materials is a low lattice thermal conductivity.
Here we present a combined neutron scattering and theoretical investigation of the lattice dynamics in the type I clathrate system 
Ba--Ge--Ni, which fulfills this requirement. We observe a strong hybridization between phonons of the Ba guest atoms and acoustic phonons of the Ge-Ni host structure over a wide region of the 
Brillouin zone which is in contrast with the frequently adopted picture of isolated Ba atoms in Ge-Ni host cages. It occurs without a strong decrease of the acoustic phonon lifetime which contradicts 
the usual assumption of strong anharmonic phonon--phonon scattering processes. Within the framework of ab--intio density functional theory calculations we interpret these hybridizations as a series of anti-crossings 
which act as a low pass filter, preventing the propagation of acoustic phonons. To highlight the effect of such a phononic low pass filter on the thermal transport, we compute the
contribution of acoustic phonons to the thermal conductivity of Ba$_8$Ge$_{40}$Ni$_{6}$ and compare it to those of pure Ge and a Ge$_{46}$ empty-cage model system.
\end{abstract}


\maketitle


A central issue in thermoelectrics research is to find materials that generate a high electromotive force under a small applied thermal gradient. The best thermoelectric materials (TE) should therefore combine a low 
lattice thermal conductivity with high electrical conductivity and thermopower. An efficient and economic way to reduce the thermal conductivity of bulk TE materials, without degrading their electronic properties, 
is to take advantage of inelastic resonant scattering between heat-carrying acoustic and non-propagative phonons, arising from isolated impurities with internal oscillator degrees of freedom. The librational vibration of molecules 
incorporated in crystal structures is a well-known example of such 
scattering centers and was formerly observed in KCl ionic crystals, doped with anionic molecules \cite{Walton-PRL74}, or in the filled water nanocages of clathrate hydrates \cite{Chazallon-PCCP2002,Gutt-JCP2002}. 
In TE clathrates and skutterudites, in which the phonon dispersions were recently investigated by means of inelastic neutrons scattering (INS) \cite{CHLee-JPConf07,Christensen-Nature08,CHLee-JPSJ2006,Yang-PCM2007}, the localized 
resonators are so-called rattling phonon modes of loosely bonded guest atoms in oversized atomic cages. The effect of inelastic resonant scattering \cite{Pohl-PRL1962,Klein-PR1969} was described theoretically, using 
the relaxation-time approximation of the Boltzmann equation, in which the lattice thermal conductivity, $\kappa_L$, for a cubic crystal can be expressed as 

\begin{equation}
\kappa_L=1/3\int_0^{\omega_{max}} C_v(\omega) v(\omega)^2 \tau(\omega) \rho_0(\omega) d\omega\quad.
\label{eq:Kappa}
\end{equation}

\noindent Here $C_v(\omega)$ is the phonon specific heat per unit volume and unit frequency, $v(\omega)$ is the mean group velocity for phonon frequency $\omega$, $\tau(\omega)$ is the mean time between collisions that destroy 
the heat current for all phonons of frequency $\omega$ and  $\rho_0(\omega)$ is the normalized vibrational density of states. 
The most common approach to evaluate the scattering of acoustic phonons by a resonator with frequency $\omega_0$ is to add a phenomenological relaxation rate, $\tau_{R}^{-1}(\omega)=C\frac{\omega^2}{(\omega^2-\omega_0^2)^2}$, 
where $C$ is a constant proportional to the concentration of the resonant defects and to the strength of the resonator-lattice coupling. This has been used to fit experimental data for clathrates and 
skutterudites \cite{Yang-PRB2003,Cohn-PRL1999}. At a microscopic level, this implies the existence of a coupling and an anti-crossing behavior between the non-dispersive phonons of the isolated resonators and the acoustic phonon 
branches leading to the opening of a gap surrounding $\omega_0$ in the dispersion of the acoustic branch \cite{Toberer-JMC2011}. INS measurements on the clathrate Ba$_8$Ge$_{30}$Ga$_{16}$ \cite{CHLee-JPConf07,Christensen-Nature08} and the skutterudite 
CeRu$_4$Sb$_{12}$ \cite{CHLee-JPSJ2006} evidence such an anti-crossing. 
However, in these previous studies, the authors conclude that the remarkably low $\kappa$ values of these materials cannot fully be ascribed to the effect of guest vibrations on lifetime $\tau(\omega)$ 
or group velocity $v(\omega)$ of the acoustic phonon modes (see also \cite{Toberer-JMC2011}). On the other hand, the picture of isolated guest atoms in a host cage and the microscopic mechanism responsible for the scattering of heat carriers 
were revisited by INS studies of the phonon dynamics in polycrystals of the skutterudites La/CeFe$_4$Sb$_{12}$ \cite{Koza-Nature2008}. 
The authors demonstrated that the guest atoms are coherently coupled with the host-lattice dynamics and associated with 
low energy optical phonon modes which are characterized by a small group velocity. It was then proposed that these low frequency phonon branches open new paths for the occurrence of Umklapp scattering of heat-carrying acoustic phonons,
that is, the backscattering of heat carriers by an elastic or a multiphonon process mediated by a lattice vector \cite{Koza-Nature2008,CHLee-JPSJ2006,Christensen-Nature08}. This
mechanism obeys momentum and energy preservation; hence, it is in contrast to inelastic resonant scattering associated with energy dissipation at isolated atoms. The reason to invoke such ph-ph processes in complex unit cell materials is the presence of flat optical phonon modes at low energies which are expected to enhance ph-ph scattering processes. However, to date experimental evidence for this claim is missing, - it would require measuring the corresponding broadening of the energy width of the phonons involved. Yet, the available INS measurements on single 
crystals \cite{Christensen-Nature08,CHLee-JPSJ2006} as well as our data demonstrate that the finite lifetime of acoustic phonons is not small enough to explain such a low $\kappa$, meaning that no fingerprints of 
strong Umklapp scattering could be evidenced. Thus, the microscopic origin for the low $\kappa$ remains elusive. 
In particular the respective roles played by guest atoms and unit cell complexity are ambiguous \cite{Macia-CMA2011}.\\

In this paper, we present in section I the experimental mapping of the transverse acoustic (TA) and the guest phonon dispersions in a single crystal of the clathrate 
Ba$_{8}$Ge$_{42.1}$Ni$_{3.5}\square_{0.4}$ \cite{Nguyen-Dalton2010} by INS, covering the entire first Brillouin zone (BZ) and a wide energy range. The high quality of 
the data allows us to analyse the relative change in intensity of acoustic and optical phonons. We evidence a gapless dispersion of the TA phonons in absence 
of a strong broadening of their energy widths when approaching the energy of the optical phonons. Moreover the existence of a mechanism of spectral weight transfer between 
acoustic and optical phonons resulting from strong hybridizations is visible over a wide region in reciprocal space. In section II, we interpret our experimental results 
in the framework of ab--intio DFT calculations comparing the lattice dynamics of pure Ge, the empty Ge$_{46}$ framework structure and the Ba$_{8}$Ge$_{40}$Ni$_{6}$ clathrate. 
By a direct comparison with the experimental findings, we highlight the roles played by unit cell complexity and guest atoms for the lattice dynamics, respectively. 
We conclude that the low lattice thermal conductivity is due to a phononic filter effect.


\section{I. Inelastic neutron scattering}
\label{sI}

Subsection \textit{a)} of this paragraph, contains details on the investigated sample, while in subsections \textit{b)} and \textit{c)} we provide the basic 
theoretical background of phonon measurements by INS and explain how our study differs from previous measurements done on similar materials. In subsection \textit{d)} 
we report our results together with our analysis and interpretation of the data. Finally, in subsection \textit{e)} we discuss technical details with respect to 
the validity and reliability of our data analysis.

\subsection{a) Sample preparation and physical properties}
\label{ss:a}

A single crystal of the clathrate Ba$_{8}$Ni$_{3.5}$Ge$_{42.1}\square_{0.4}$ (space group Pm$\bar{3}$n, $a$ = 10.798(2) \AA , \o = 8 mm and height= 30 mm), as shown in Fig.~\ref{fig1} a), was grown from the melt, using the Bridgman 
technique. A detailed analysis of its crystal structure and its thermoelectric properties were reported in \cite{Nguyen-Dalton2010}. This study revealed Ba$_{8}$Ni$_{3.5}$Ge$_{42.1}\square_{0.4}$ 
to be a thermoelectric (TE) n-type metal with a relevant value for the dimensionless TE figure of merit $ZT$ at high temperatures, thanks to its very low thermal conductivity of about $\kappa\sim~1.2\ $W/mK at 300K.

\subsection{b) Neutron intensity, phonon dynamical structure factor}
\label{ss:b}

\begin{figure}[t]
  \centering \includegraphics[width=0.35\textwidth]{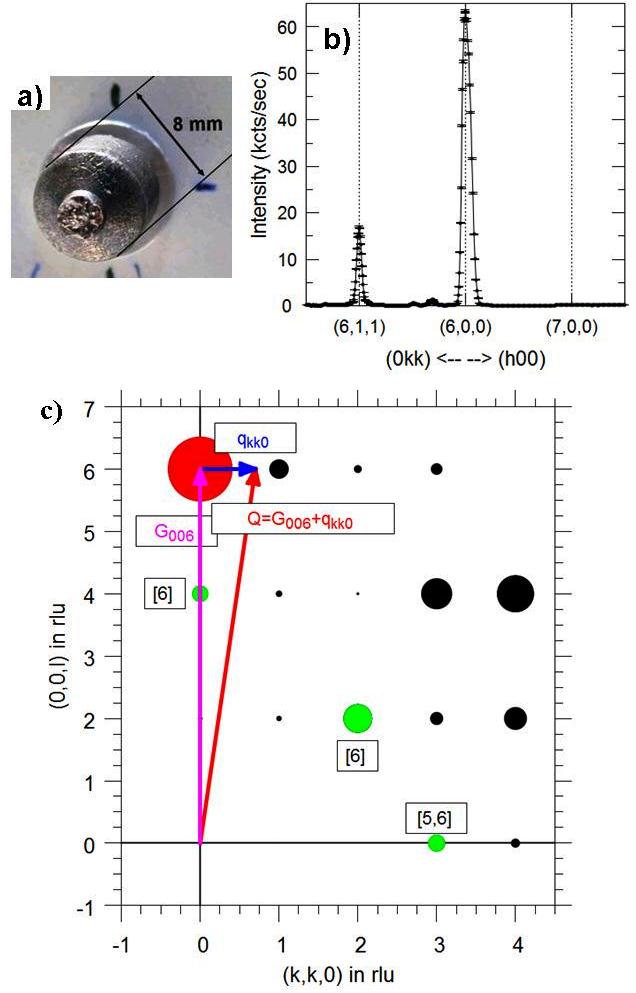}
  \caption{ \textit{(Color online)} {\bf a)}: The measured single crystal of Ba$_{8}$Ni$_{3.5}$Ge$_{42.1}\square_{0.4}$ \cite{Nguyen-Dalton2010} {\bf b)}: Elastic neutron scans around the Bragg peak $(600)$ along the transverse, 
  $[110]$, and longitudinal, $[100]$, directions. {\bf c)}: Bragg peak intensities in the $\{[100];[011]\}$ plane, calculated for a neutron wave vector $k=2.662$\AA$^{-1}$. The pointsize is proportional to the nuclear elastic 
  structure factor. The intensity of the strongest Bragg peak $(600)$ is then compared to those used in \cite{CHLee-JPConf07} and \cite{Christensen-Nature08}. As sketched and explained in subsection \textit{I.b)}, INS measurements of TA phonons are obtained by scanning the reduced wave vector, $q$, along the $[011]$ direction from the Bragg peak $(600)$.   } 
\label{fig1}

\end{figure}

In case of coherent inelastic nuclear scattering by a phonon of branch $j$, with energy $\omega_{q,j}$ and polarization vector $\xi^j_{\omega_{q,j}}$, the neutron diffusion function is written as: 
\begin{equation}
S_{ph}(\vec{Q},\omega) = n(\omega) \frac{\vert F_D^j(\vec{Q})\vert^2}{\omega_{q,j}} \delta (\omega-\omega_{q,j}) \delta (\vec{Q}-\vec{q}-\vec{G})
\label{eq:INS2}
\end{equation}
\noindent where $n(\omega)=\frac{1}{1-exp(-\hbar\omega/k_B T)}$ is the Bose factor, $\vec{Q}=\vec{q}+\vec{G}$ is the scattering vector given by the nearest reciprocal lattice vector $\vec{G}$ and
the phonon wave vector $\vec{q}$. Finally $F_D^j(\vec{Q})$ is the dynamical structure factor (DSF) defined as 
\begin{equation}
F_D^j(\vec{Q}) = \sum_i e^{-W_i(\vec{Q})} \frac{b_i}{\sqrt{M_i}} e^{i\vec{Q}.\vec{r}_i} \lbrace \vec{Q}.\vec{\xi}_j^i(\vec{Q}) \rbrace
\label{eq:INS3}
\end{equation}
\noindent where $b_i$, $\vec{r}_i$, $M_i$ and $W_i (\vec{Q})$ are coherent scattering length, fractional coordinates, mass and Debye-Waller factor of the i'th element, respectively. 
The scalar product $\lbrace \vec{Q}.\vec{\xi}_j^i(\vec{Q}) \rbrace$ contains the phonon polarization and can thus be used to distinguish longitudinal and transversal phonon modes by choosing the
the appropriate combination of phonon wave vector $\vec{q}$ and reciprocal lattice vector $\vec{G}$. The polarization vectors of longitudinal and transversal phonons are parallel and perpendicular to the phonon wave vector $\vec{q}$,
respectively. In our case, as sketched in Fig.\,\ref{fig1} c), measurements were conducted around the $\vec{G}=(006)$ reflection, scanning along $\vec{q}_{[110]}$. 
In this setting longitudinal phonons, polarized parallel to $\vec{q}$, are perpendicular to $\vec{G}$ and therefore only transverse components are selected.\\
It is then worth to denote the diffusion function in the long wavelength limit, for $\vert\vec{q}\vert<<\vert\vec{Q}\vert$. It can be shown that for acoustic phonon modes, close to the Brillouin zone center, all atoms are 
vibrating in phase and with equal displacements $\xi_j^i(\vec{Q})/\sqrt{M_i}$. Consequently, for $T>>\omega$, the diffusion function, Eq.~\ref{eq:INS2}, 
becomes proportional to the nuclear structure factor (i.e. the Bragg intensity): 
\begin{equation}
 S_{ph}(\vec{Q},\omega) \varpropto {G^2}\frac{\vert F_B(\vec{Q})\vert^2}{\omega_j^2(\vec{q})}
\label{eq:INS4}
\end{equation}
\noindent 
Thus, the intensity is proportional to the nuclear structure factor of the respective Bragg peak.  Moreover, close to the Brillouin zone center and for pure acoustic phonons $S_{ph}(\vec{Q},\omega)\times\omega_j^2(\vec{q})$ is constant. 

\subsection{c) Measurement strategy and data analysis }
\label{ss:c}

\begin{figure}[t]
  \centering \includegraphics[width=\columnwidth]{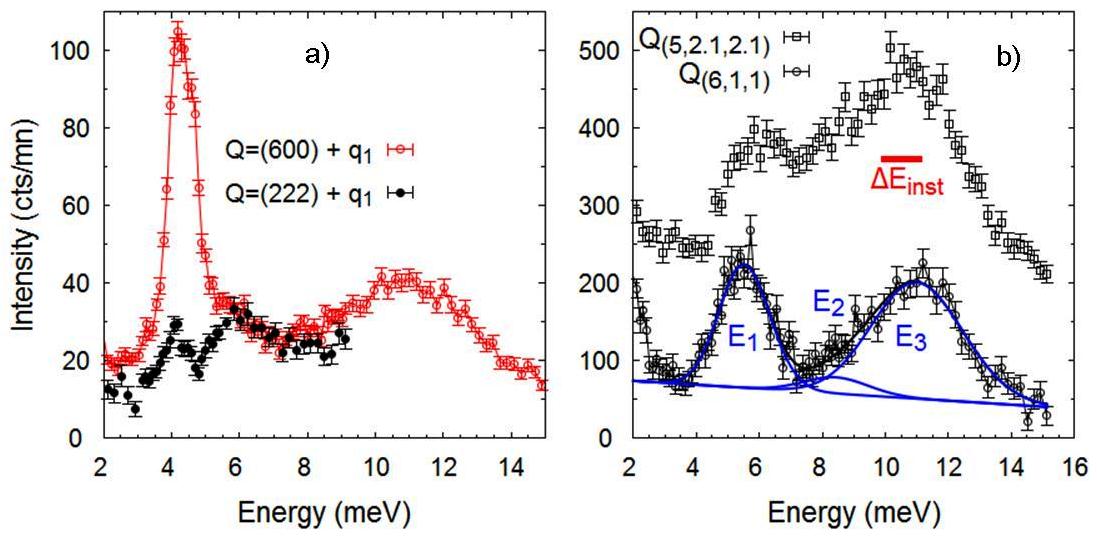}
  \caption{\textit{(Color online)} {\bf a)}: Energy scans close to the Bragg peaks Q=$(600)$ or Q=$(222)$, at wave vector $q_{1}=0.261\AA^{-1}$ in transverse geometry. {\bf b)}: Energy scans measured at the constant wave 
  vectors $\vec{Q}=(611)$, open circles, and $\vec{Q}=(5,2.1,2.1)$, open squares. 
Three Gaussian profiles (blue lines) centered at the energies labeled by $E_1$, $E_2$ and $E_3$ were used to fit the total phonons line shape (black line). The instrumental energy resolution, $\Delta E_{inst}$=1.1 meV, 
is indicated by the red line}
\label{fig2}
\end{figure}
INS experiments were carried out on the thermal triple axis spectrometer 2T at the Orph\'ee reactor of the Laboratoire L\'eon Brillouin (LLB, France). All measurements were conducted at room temperature in 
fixed $Q$ mode with a final wave vector of 2.662 \AA$^{-1}$. To eliminate higher-order harmonics graphite filters were used. The sample was mounted in a vacuum chamber and aligned in the plane $\{[100];[011]\}$ such that wave vectors 
of the form Q=2$\pi/a(hkk)$ were accessible as depicted in Fig.~\ref{fig1}.c. Wide elastics scans around the Bragg peak $(600)$ are shown in the Fig.~\ref{fig1}. The bulk mosaicity is of 0.75$^{\circ}$.\\ 
In the long wavelenght limit the acoustic phonon intensity is directly proportional to the Bragg intensity as outlined above, see Eq.~\ref{eq:INS4}. Therefore acoustic phonons have to be measured near strong nuclear Bragg reflections.
The main technical difference between the present and previous INS experiments on clathrates is that we could access a higher $Q$-range and thus measurements around the strongest Bragg peaks
could be conducted. As sketched in Fig.~\ref{fig1} b, our measurement was conducted around the strongest reflection $\vec{G}=(600) $, following the transverse $[0kk]$ direction, while previous INS studies of transverse
phonon branches were carried out around less intense Bragg peaks $(400)$, $(222)$ and $(330)$ \cite{Christensen-Nature08,CHLee-JPConf07}. As a consequence of the INS cross section detailed above, the TA phonons in our data appear
much stronger than the guest phonon modes and with a significantly 
enhanced resolution, when compared to the results in Refs. \cite{CHLee-JPConf07} and \cite{Christensen-Nature08}.  To emphasize this aspect, we compare in Fig.~\ref{fig2} a) transverse phonon spectra measured at the same wave vector 
$\vert \vec{q}_1\vert$($0.261\AA^{-1}$) either around the Bragg peaks Q=$(600)$ or Q=$(222)$. Another consequence of the measurement around the Bragg peak $\vec{G}=(600) $, which follows directly from energy and momentum conservation laws, 
is that our 
measurements go up to 14 meV while data in Ref. \cite{Christensen-Nature08} stop at 7 meV. This is an important advantage which allows us to follow the spectral weight distribution and transfer among the different acoustic and optic 
phonon branches. \\
Thanks to the high data quality, we are then also able to perform an advanced and reliable analysis in order to extract properly the intrinsic energy width, the energy position and the dynamical structure factor, or norm, of the 
contributing phonons. In fact, realistic INS measurements on a conventional (mono-detector) neutron triple axis spectrometer mean scanning the ($\vec{Q},\omega$) space step by step with a resolution function. The neutron intensity at 
a given point $(\vec{Q}_0,\omega_0)$ is given by the convolution product of this instrumental resolution function, 
$R(\vec{Q},\omega)$, and the diffusion function, $S(\vec{Q},\omega)$, which contains all the information of the dynamics in the sample probed: 
\begin{equation}
I(\vec{Q}_0,\omega_0) = S(\vec{Q},\omega) \otimes R(\vec{Q}-\vec{Q}_0,\omega-\omega_0)
\label{eq:INS1}
\end{equation}   
\noindent 

The resolution function is a 4-dimensional function which couples the energy and the reciprocal space. A cut of this resolution function in the \{$\vec{q}_{0kk},\omega$\} plane is shown in Fig. \ref{fig5} a). 
As one can see, the size of this function is not negligible with regard to the phonon branch dispersion. Therefore, in order to extract properly the intrinsic position, width and DSF of a phonon from the measured scattering 
profile, we use a fitting procedure which numerically convolutes a model phonon cross section with the spectrometer resolution. It enables us to take into account the local curvature of the phonon branch over the range of the
instrumental resolution, i.e. the fact that the energy resolution for an acoustic phonon mode is not  independent of its slope. Only this method allows for a reliable extraction of the characteristics of phonons from INS data. 
The DSF has so far not been discussed in any neutron work on clathrates. (For a similar analysis on the skutterudite CeRu4Sb13 we refer to \cite{CHLee-JPSJ2006}).\\

\subsection{ (d) INS Results }
\label{ss:b}

\begin{figure}[t]
  \centering \includegraphics[width=\columnwidth, height=17.5 cm]{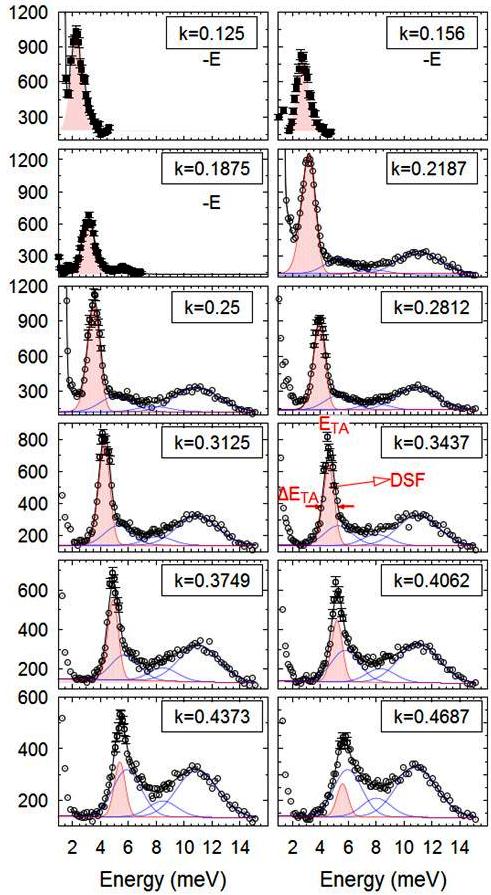}
  \caption{\textit{(Color online)} Energy scans with constant $\mathbf{Q}=(600)+\mathbf{q}(0kk)$ in the first BZ, $k<$0.5. The solid black lines are fits as explained in the text. Red lines and filled areas depict contributions of 
  TA phonons while blue lines are representatives of the peaks at $E_{1}$, $E_{2}$ and $E_{3}$. For scans labeled by -E the negative energy region is shown (Stokes side). }
\label{fig3}
\end{figure}

\begin{figure}[t]
  \centering \includegraphics[width=\columnwidth, height=13 cm]{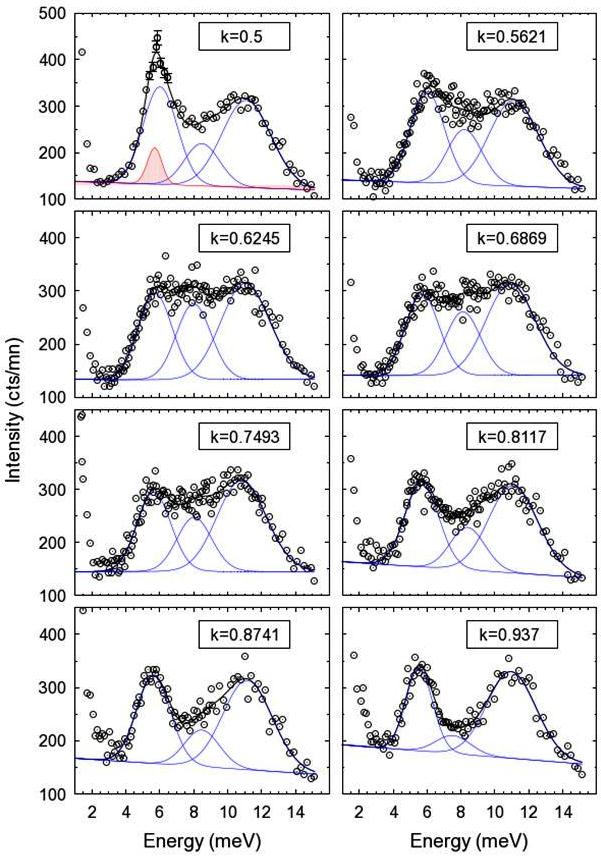}
  \caption{\textit{(Color online)} Energy scans with constant $\mathbf{Q}=(600)+\mathbf{q}(0kk)$ for $k\geq 0.5$. The solid black lines are fits as explained in the text. Red lines and filled areas depict contributions of TA phonons 
  while blue lines are representatives of the peaks at $E_{1}$, $E_{2}$ and $E_{3}$.}
\label{fig4}
\end{figure}

\begin{figure}[t]
\centering \includegraphics[width=\columnwidth]{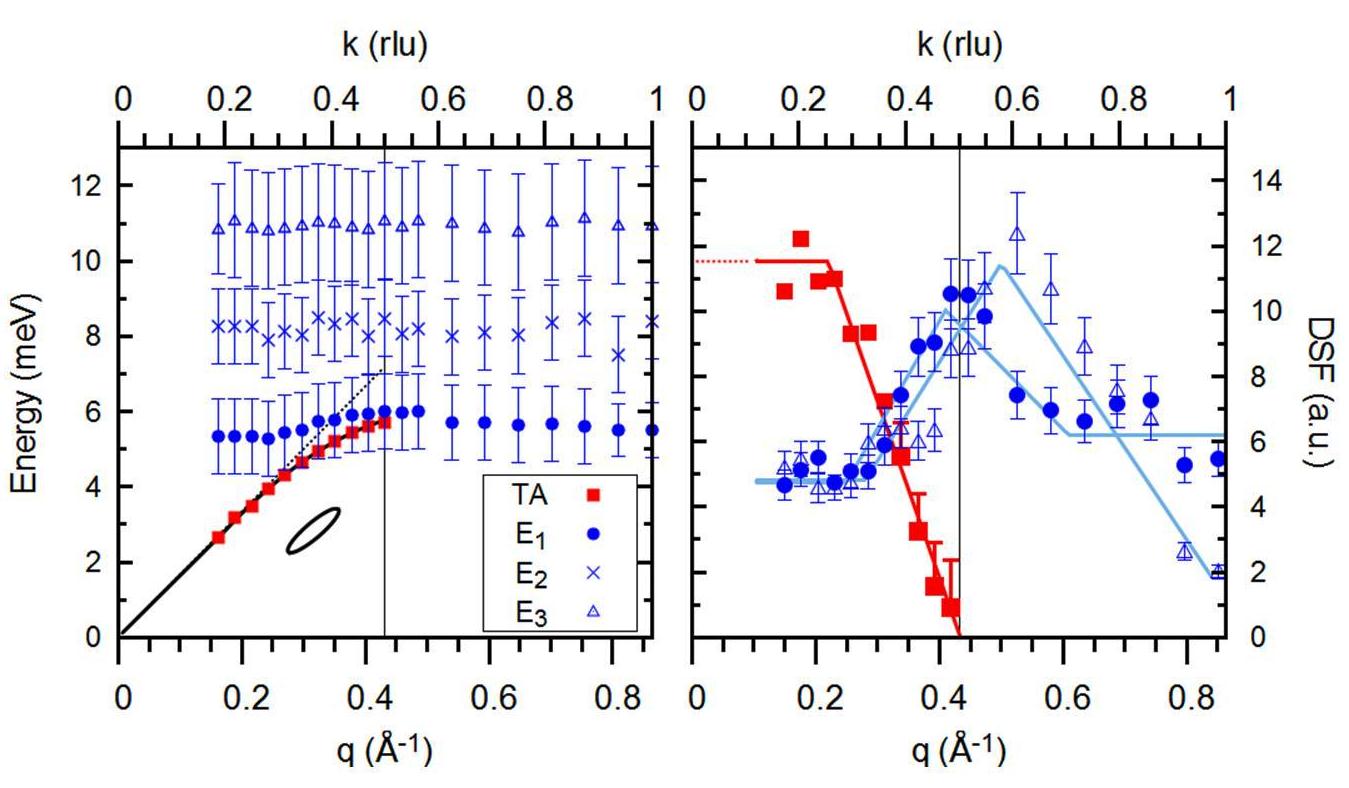}
\caption{\textit{(Color online)} {\bf{a)}} Energy dispersions of the TA and the three optic-like ($E_{1}$, $E_{2}$ and $E_{3}$) phonons. The size of the error bars 
represent the phonon linewidth (FWHM). 
The ellipse is a cut of the instrumental resolution volume (axis sizes: 0.03 \AA$^{-1}$ (0.034 rlu) along the $(0kk)$ axis and of 0.24 meV along the energy axis). {\bf{b)}} DSF 
as function of $q$ in the direction $(0kk)$.}
\label{fig5}
\end{figure}

\begin{figure}[t]
  \centering \includegraphics[width=\columnwidth]{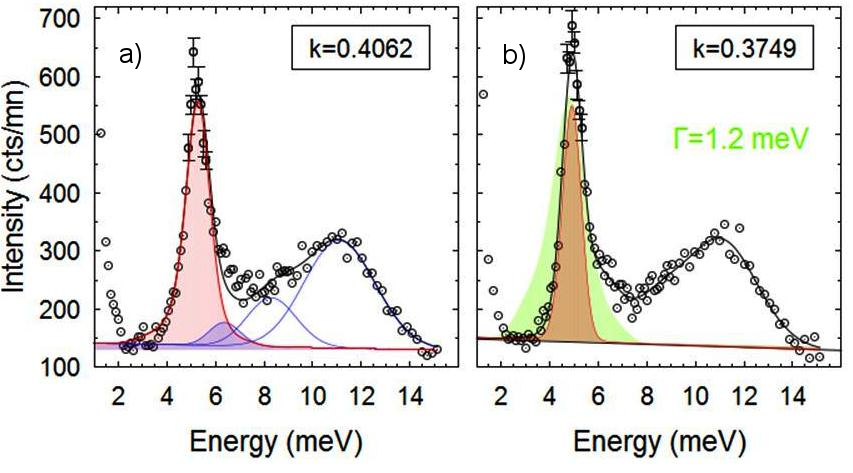}
  \caption{\textit{(Color online)} {\bf{a)}} Fit to the scan at $k$=0.4062 rlu, obtained by forcing a reduction of the intensity of the optic phonon mode at $E_{1}$ (shaded in blue). 
The fits provide the upper limit of the intensity of the acoustic phonon as indicated by error bars in Fig. \ref{fig5} a). {\bf{b)}} green area represents the simulated profile of the phonon measured at $k$=0.3749 rlu with a finite energy width of $\Gamma=1.2$ meV. }
\label{fig6}
\end{figure}

The energy scans, shown in Fig.~\ref{fig2} b), were performed far from any acoustic phonon branch at the wave vectors $(611)$ and $(5, 2.1, 2.1)$  to localize the low lying optical phonon modes. 
The energy spectra at $(611)$ were analyzed by using three Gaussian profiles centered at energies of $E_{1}$=5.5(0) meV, $E_{2}$=8.2(0) meV and $E_{3}$=10.8(5) meV with full width at half maximum (FWHM) 
values of $\Delta E_1\sim 2$ meV, $\Delta E_2\sim 2$ meV and $\Delta E_3\sim 3$ meV. Since the linewidth of these peaks is larger than the energy resolution, several phonon branches can be contained in a peak. Later, the peaks $E_{1}$ and $E_{2}$ will be identified as stemming from localized 
Ba motions in the large Ba(Ge,Ni)$_{24}$ cages, while the peak $E_{3}$ contains both optical phonon modes from the (Ge,Ni)$_{46}$ host framework and guest phonon modes, in agreement with previous INS 
\cite{CHLee-JPConf07,Christensen-PhysicaB06,Christensen-Nature08,Koza-PRB2010,Johnsen-PRB10}, Raman \cite{Takasu-PRB2010} and theoretical \cite{Dong-JAP2000,Koza-PRB2010} reports.\\ 
In a next step the TA phonon branch was investigated by performing energy scans around the Bragg peak $(600)$ for a set of reduced phonon wave vectors, $\mathbf{q}=(0kk)$, 
covering an entire BZ ($0\leq k \leq 1$). 
Raw data are shown in Fig.\,\ref{fig3} for $k<$0.5 and in Fig.\,\ref{fig4} for $k\geq$0.5. The model phonon cross section, applied for our fitting procedure, 
consists of a damped harmonic oscillator for TA phonons and three Gaussian functions (centered at energies $E_{1}$, $E_{2}$ and $E_{3}$)
for the low-lying optical phonons as beforehand identified. In Fig.\,\ref{fig2} red lines and shaded areas depict the contributions of the TA phonons, while the blue lines represent the three optic-like 
excitations at $E_{1}$, $E_{2}$ and $E_{3}$. The deconvoluted energy positions and dynamical structure factors (DSF) of both the TA phonons and the three optic-like excitations are 
shown in Fig.\,\ref{fig5} as functions of $q$.\\
For $k<$0.5, the phonon spectra consist of a main peak, stemming from the TA phonon branch (red lines in Fig.\,\ref{fig3}), which appears at energies distinctly lower than the optical excitations. 
As shown in Fig.\,\ref{fig5} a), the energy positions of these peaks describe a dispersive branch below the optical continuum. This branch does not cross any optical phonon mode and no gap is observed following its 
dispersion. For wave vectors below about $k$=0.25, the energy width of the TA modes is resolution limited and their DSF fluctuates around a constant value. 
In the limit $\omega\rightarrow 0$, the slope of the TA phonon branch yields the transverse sound velocity $\nu_t^{(011)}$=2530 m/s, comparable to that one of Ba$_8$Ge$_{30}$Ga$_{16}$ 
\cite{Christensen-Nature08}. The dispersion of the TA mode deviates from linearity for energies above about 4 meV ($\sim$ 46 K). For $0.25 \le k \le 0.4$, 
although there are still well defined TA phonons with resolution limited energy widths, one observes an abrupt decrease of their DSF. The DSF of TA phonons drops by more than 80\% between $k=0.25$ and $k=0.4$, as seen 
in Fig.\,\ref{fig5} b). The analysis of the experimental data becomes more delicate for wave vectors above $k$=0.4 where intensities and energy positions of the TA branch and the low lying optical excitations 
come closer. However, as can be observed in the raw data in Fig.\,\ref{fig3} as well as in Fig.\,\ref{fig6} where only the phonon spectrum measured at k=0.4062 is depicted, there are always well defined TA phonon peaks 
located at energies below the optical excitations. The profile of these peaks is continuous until a break at about the energy E$_1$ of the low lying optical excitation. 
While the TA phonon branch fades away, the intensities of the guest phonon peaks at $E_1$ and $E_2$ change (see Fig.\,\ref{fig5} b)). The DSF of the guest peak at $E_1$ 
increases from a constant value for $k\leq$0.25, reaches its maximum at $q$=0.43 \AA$^{-1}$ ($k$=0.5) and then returns to roughly the same constant value.

A similar $q$-dependence of the DSF is observed for the guest peak at $E_2$ with the maximum slightly shifted to a higher $q$ of about 0.54 \AA$^{-1}$ ($k$=0.62). For the peak at $E_3$ no sizable change 
is observed over the entire $q$-range. Note that the qualitative changes of the guest DSFs are clearly 
visible on the raw data in Figs.\,\ref{fig3} and \ref{fig4} (compare e.g. data measured at $k$=0.34 to those obtained at $k$=0.46, $k$=0.62 and $k$=0.87). As evidenced in Fig.\,\ref{fig5} b), the $q$-dependence of the DSF 
seems to arise from a transfer of spectral weight between TA phonons and guest vibrations. As $k$ increases, the spectral weight of the TA phonons is first transferred to the first 
guest phonon peak at $E_1$ and then to the second one at $E_2$. Interestingly, similar effects have been observed in quasicrystals and large unit cell crystals \cite{Takeuchi-PRB2006,Boissieu-Nature2007}.
The energy of the guest phonons at $E_1$ increases significantly during this process, shifting from 5.4 meV at the zone center to 6 meV at $k$=0.5 without any sizable change in its energy width (see Fig.\,\ref{fig5} a)). \\
To conclude, firstly, our data put in evidence a continuous dispersion of the TA phonon branch throughout the whole BZ zone with a rapid bending together with a drastic decrease of the spectral weight as $k$ increases beyond $0.25$ or 
for energies higher than 4 meV, as summarized in Fig.\,\ref{fig5} a) and b). The most surprising and relevant effect observed is precisely that the spectral weight of TA phonons decreases abruptly without a strong broadening of their 
energy profiles. In others words, within our resolution, no finite energy damping or finite lifetime is needed to fit the TA phonon profiles. 
Note that the lack of line broadening in the deconvoluted data is not meant to demonstrate the complete lack of anharmonic phonon scattering processes, it only emphasizes that anharmonic effects are by far not 
as big as assumed in previous studies. Therefore it becomes evident that it is not correct to 
consider a strong decrease of the relaxation time $\tau(\omega)$ as function of energy in Eq. \,\ref{eq:Kappa} to account for the effect of guest phonon modes. Moreover, the TA phonon branch disperses continously without any 
fingerprint of a gap induced by a crossing with the low lying optical phonon branch in contrast to what is reported by \cite{Christensen-Nature08}. This makes it obvious that adding resonant like relaxation times in Eq.\,\ref{eq:Kappa} 
to account for the effect of vibrations of the guest atoms is incorrect. \\
Secondly, our data highlight a mechanism of spectral weight transfer between TA and guest phonon modes which occurs over a wide region of the BZ, from $k$=0.25 to $k$=0.5. Such transfer of spectral weight was observed in the skutterudite CeRu$_4$Sb$_{12}$ \cite{CHLee-JPSJ2006}.
Note that along a given direction across the BZ, neither an intensity change nor an energy dispersion would be expected for guest phonons, if they were assumed to be isolated oscillators.
Since we observe changes of the spectral weight of both  $E_1$ and $E_2$ guest phonons as well as a dispersion of the peak $E_1$, the picture of a "freely rattling" guest atom in a host cage is not applicable in the case of clathrates, 
as previously concluded from INS experiments on skutterudites \cite{Koza-Nature2008}.

\subsection{ (e) Validity and reliability of the data analysis  }
\label{ss:c}

Crucial for our data analysis is the ability to differentiate between the acoustic and optical parts of the measured spectra even in the region where both 
contributions are superimposed, \textit{i.e.} for $k\geq$0.4. The TA phonons have lost already 80 $\%$ of spectral weigth at $k=0.4$, see Fig.\,\ref{fig5} b), with no evidence of any broadening in energy. 
Thus the main effect of spectral weigth transfer between the TA and optical phonons occurs then well below $k=$0.4 where the precision of the fits is excellent. We now discuss the data analysis of the energy scan measured at $k$=0.4062, 
shown in Fig. \ref{fig6}, in which the distance in energy between the TA phonon and the optical phonon at $E_1$ is less than 1 meV. In our data analysis, the surprising result is that, 
within our resolution, no finite energy width is needed to fit the profile of the TA phonons. The observed energy width arises simply from the integration of all the phonon modes 
at $0.4062\pm\delta k$ contained in the instrumental resolution function. However, one can try to force the occurrence of a finite energy width by forcing the intensity of the 
$E_1$ peak contribution to decrease as compared to the fitting model proposed in Fig.\,\ref{fig3}. The result, shown in 
Fig.\,\ref{fig6} a), clearly yields a poorer fit with an intrinsic energy width of 0.45 meV and an increase of about 25\% of the spectral weigth of the TA mode. 
The resulting limits for the TA phonon intensity are indicated by the red line on the y-axis in Fig.\,\ref{fig5} b). We thus clearly evidence that, independtly of the 
fitting model, the acoustic phonon looses abruptly its intensity when approaching $E_1$. \\ 
As demonstrated above, the upper limit for the TA phonon energy width is roughly 0.45 meV. In a simple Debye approach of the thermal conductivity based on 
Eq. \,\ref{eq:Kappa} and on our measured TA phonons dispersion (see below), we estimate that the relaxation time that would be required to reproduce the low experimental 
value of the thermal conductivity in our material is of about $\tau\sim 0.2$ ps corresponding to a phonon energy width of about 3.3 meV. In Fig.\,\ref{fig6} b), 
we simulate the profile of the TA phonon measured at $k$=0.3749 with a finite energy  width of 1.2 meV, dashed line, 
and we compare it with the result of our best fit. It evidences that the main microscopic mechanism responsible for the low thermal conductivity cannot be a large relaxation time.


\section{II. DFT CALCULATIONS}

\begin{figure}[t]
  \centering \includegraphics[width=0.35\columnwidth,angle=270]{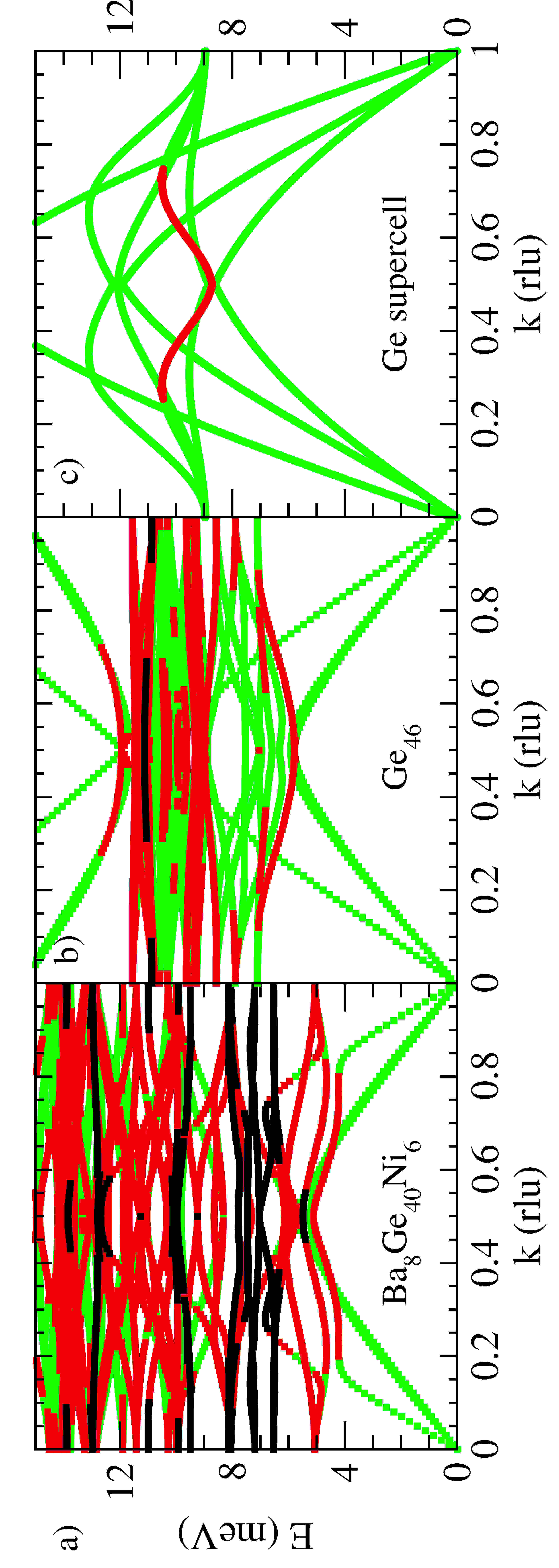}
  \caption{\textit{(Color online)} Calculated dispersion curves along the reciprocal space direction $(6kk)$ for a) Ba$_8$Ge$_{40}$Ni$_6$, b) Ge$_{46}$ and c) diamond Ge. Modes are color coded with respect to their PR. 
While green means a high PR, red and black stand for intermediate and low PR, respectively (see text). The Energy axis of panel a) is rescaled by a factor 1.4 to match the experimental data.}
\label{fig7}
\end{figure}

To deepen our understanding of the INS measurements and of the coupling mechanism between the guest and host phonons, we have performed an ab initio density functional theory (DFT) study of the lattice dynamics and the INS cross section 
in the cubic high-symmetry configuration Ba$_8$Ge$_{40}$Ni$_{6}$ and 
an empty Ge$_{46}$ framework model system. 

Ab-initio density functional theory (DFT) calculations were conducted using the DFT code VASP \cite{Kresse-PRB1996} applying the projector augmented wave generalized gradient method (PAW-GGA). First the unit cell of Ba$_8$Ge$_{40}$Ni$_{6}$ 
was relaxed to its electronic ground state, using high precision settings with a 4$\times$4$\times$4 k-point mesh and a convergence criterion of residual forces of less than 10$^{-4}$ eV/\AA. 
After reaching the ground state, symmetry non-equivalent displacements were introduced and the resulting Hellmann-Feynman forces were calculated and used to determine the 
dynamical matrix by use of the PHONON package \cite{Parlinski-2005}. Different displacements, ranging from 0.03 \AA\, to 0.08 \AA, were investigated, yet no significant differences
in the phonon properties could be observed. The presented calculations were then conducted for displacements of 0.05 \AA.
By solving the dynamical matrix for selected $q$-values, phonon eigenfrequencies and corresponding eigenvectors can be determined within the harmonic approximation. From these eigenvectors and eigenvalues it is then possible to calculate diffusion function and DSF following Eqns. (\ref{eq:INS2}) and (\ref{eq:INS3}).

\subsection{a) Dispersion curves}
\label{ss:dispersion}

\begin{figure}[t]
  \centering \includegraphics[width=0.35\columnwidth,angle=270]{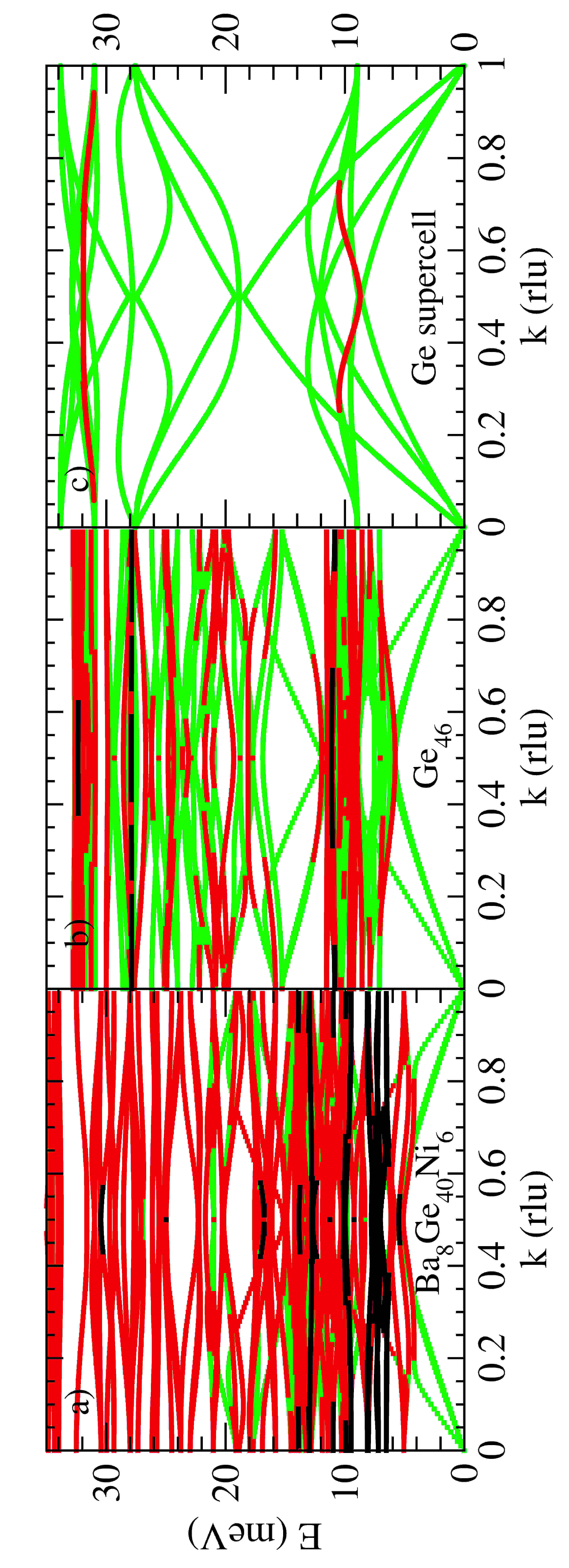}
  \caption{\textit{(Color online)} Calculated dispersion curves as in Fig. \ref{fig7}, over the full energy range.}
\label{fig7b}
\end{figure}

Figures\,\ref{fig7} and \ref{fig7b} depict a comparison of the phonon dispersion curves of Ba$_8$Ge$_{40}$Ni$_{6}$ and Ge$_{46}$ along the reciprocal space direction $[0kk]$.
As for other complex systems \cite{Euchner-PRB2011}, a rescaling of the energy axis is necessary to achieve good agreement between the absolute energy values of calculated and experimentally determined acoustic and 
low-lying optical branches in Ba$_8$Ge$_{40}$Ni$_{6}$. The rescaling parameter of 1.4 is relatively large, pointing to an underestimated Ba-Ge interaction within the DFT approach. Colors in Figs.\,\ref{fig7} and \ref{fig7b} 
represent the phonon participation ratio (PR), defined for an eigenmode $j$ as \cite{Krajci-MSE97}

\begin{equation}
 p_{j}(\omega)=\left(\sum_{i=1}^{3N}\frac{|\mathbf{\xi}_j^i(\mathbf{q})|^2}{M_{i}}\right)^2 / N\sum_{i=1}^{3N} \frac{|\mathbf{\xi}_j^i(\mathbf{q})|^4}{M_{i}^2}\quad.
\end{equation}

\noindent The PR is calculated from the phonon eigenvector $\mathbf{\xi}_{j}^i(\vec{q})$ of a given mode and the masses of the respective atoms. A PR close to 1 (green) means coherent displacements for all atoms of the same species and 
thus a collective excitation. Localized excitations, such as rattling modes, have a PR close to 0 (black), meaning that only a few atoms have large displacements, while the others are only slightly or not at all displaced.\\

The comparison of the two cage structures with the diamond Ge enables 
to disentengle different reasons for a reduction of $\kappa_l$. If we first compare Ge$_{46}$ and diamond Ge we find that many phonon modes in the complex structure are less dispersive and evidence a low PR. 
Interestingly this corresponds nicely to the results obtained in reference \cite{Dong-PRL2000}, where already the clathrate framework is found to reduce $\kappa_l$.
While neither pure Ge nore the empty Ge$_{46}$ structure evidence modes with a very low PR, the flat dispersionless branches between 6 and 8 meV in Ba$_8$Ge$_{40}$Ni$_{6}$ clearly do and can thus be identified as Ba rattling modes. 
These branches have strong impact on the acoustic phonon modes. 
First, they force the acoustic branch to bend over at lower energies, thus decreasing the group velocity of the acoustic phonons. Furthermore, they successively hybridize with the acoustic branch and lead to an important 
reduction of the PR of the 
hybridized acoustic waves, which will contribute to the dissipation of the heat current through the material. More than a simple anti-crossing, the rattling modes act as a low pass filter for propagating phonons 
which prevents propagation of heat 
carrying acoustic phonons with energies higher than the Ba rattling energies. Filtered phonons are not only dispersionless modes with small group velocities but are also characterized by a low or intermediate PR, 
meaning that only few atoms are actually involved. Unfiltered phonons are unaffected by the rattling phonons and their efficiency to propagate heat depends on the usual anharmonic scattering processes.\\

\begin{figure}[t]
  \centering \includegraphics[width=0.4\textwidth]{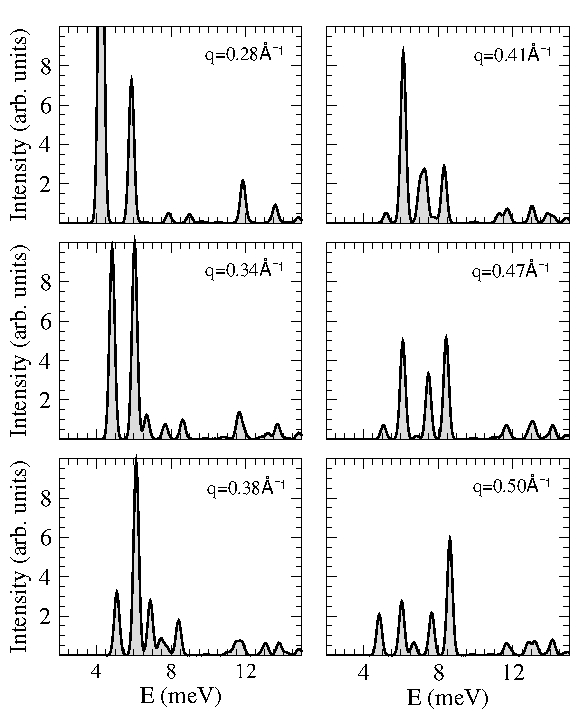} 
  \caption{Constant $q$ scans in the first BZ, determined for the same $q$ values as measured experimentally. 
The finite linewidths is obtained by convolution with a Gaussian function of 0.1 meV width.}
\label{fig8}
\end{figure}

If we now moreover investigate the simulated INS scattering profiles, a sequence of spectral weight transfer from acoustic phonons to the different rattling modes is visible, as was found for the experimental data of Ba$_8$Ge$_{40}$Ni$_{6}$, as $k$ is increased (see Fig. \ref{fig8}).
However, these effects are clearly more pronounced in the experiment, in which hybridization starts at lower $k$ values and leads to a faster decrease of the DSF of the 
TA phonon modes. This might be attributed to the fact that, in our simulation, 
we use a model which is free of disorder. Indeed it can be shown that the Ni content strongly influences the cage size and the related Ba rattling frequencies.

\subsection{b) Disorder}
\label{ss:dispersion}

\begin{figure}[ht]
  \centering \includegraphics[angle=0,width=0.45\textwidth]{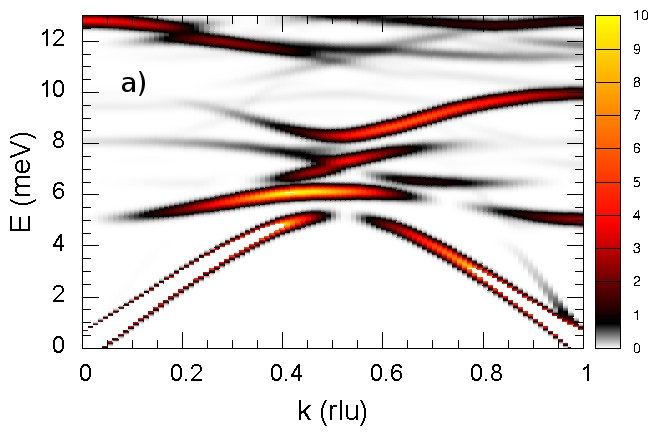}
  \centering \includegraphics[angle=0,width=0.45\textwidth]{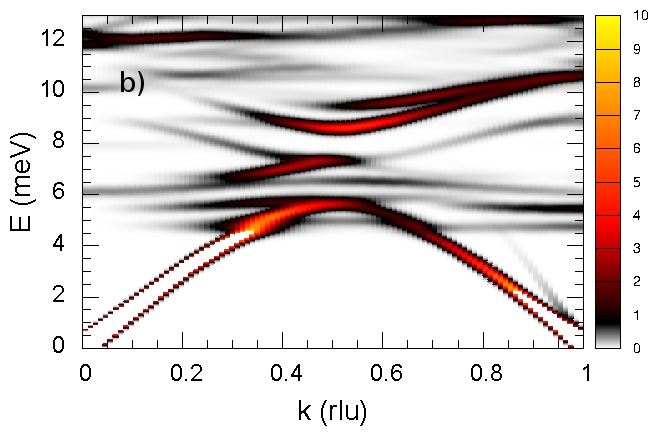}
  \caption{DSF of {\bf{a)}} Ba$_8$Ge$_{40}$Ni$_6$ and {\bf{b)}} Ba$_8$Ge$_{42}$Ni$_4$ along the direction $(6,\xi,\xi)$.}
\label{fig9}
\end{figure}

Due to the size limit which is imposed to ab-initio calculation, an investigation of disorder is rather difficult. In the Ba-Ge-Ni system, it is yet possible to create idealized structures with different 
Ni content, which still evidence high symmetry and can be described in a unit cell approach, thus enabling us to get some idea about the influence of structural disorder. The structures of Ba$_8$Ge$_{40}$Ni$_6$ 
and Ba$_8$Ge$_{42}$Ni$_4$ differ on the 6c position, which is the only site that exhibits disorder in the Ba-Ge-Ni clathrate structure. While it is fully occupied by Ni atoms in the case of Ba$_8$Ge$_{40}$Ni$_6$, 
it contains four Ni and two Ge, which are symmetrically distributed in case of our Ba$_8$Ge$_{s42}$Ni$_4$ model system.
Since the 6c position is located in the large 24-atom trapezohedron, this cage indeed is influenced by the Ni content. An increasing Ge and vacancy content on this position thus introduces distortions and breaks 
the symmetry of this cage. This symmetry--breaking is expected to broaden the frequency distribution of the encaged Ba rattling modes. And indeed by comparing the DSF of Ba$_8$Ge$_{40}$Ni$_6$
and Ba$_8$Ge$_{42}$Ni$_4$ this is evidenced (see Fig.~\ref{fig9}).
The low lying Ba rattling mode which is seen on the top panel of Fig.~\ref{fig9} is splitted in two branches on the bottom panel, thus confirming that disorder yields a broadening of the rattling modes.

\subsection{c) Phononic filter effect}

\begin{figure}[t]
 \centering \includegraphics[angle=270,width=0.9\columnwidth]{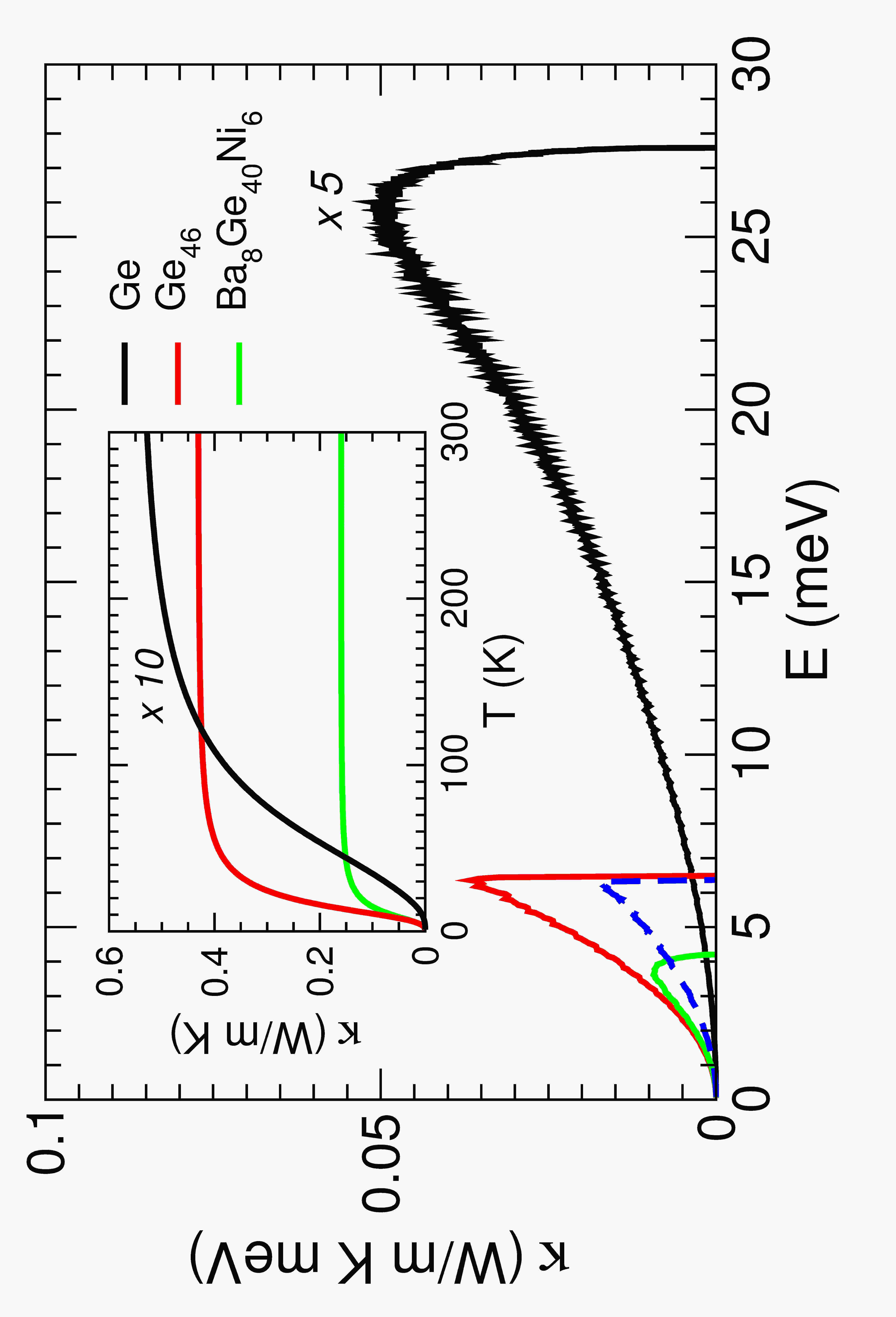}
\caption{\textit{(Color online)} Calculated frequency dependence of the lattice thermal conductivity per unit frequency for pure Ge, Ge$_{46}$ and Ba$_8$Ge$_{40}$Ni$_{6}$ at room temperature along a LA branch of the dispersion curve. The 
dashed blue line shows $\kappa_{LA}(\omega)$ for pure Ge with the same cutoff that is applied for Ge$_{46}$.
The inset shows the resulting $\kappa(T)$ for the three systems obtained from the average over TA and LA branches of the $(0kk)$ direction. 
The curves for Ge (black) are scaled by a factor 0.2 (main panel) and 0.1 (inset) for better readability.}
 \label{fig10}
\end{figure}

To highlight the effect of such a phononic low pass 
filter on the thermal transport, we compute the contribution of LA phonons, propagating along the $[011]$ direction, to the thermal conductivity of pure Ge, Ge$_{46}$ and Ba$_8$Ge$_{40}$Ni$_{6}$. From our 
ab--initio calculations for these three systems, we extract $\rho_0(\omega)$ and $v(\omega)$ for the chosen phonon branch. We then evaluate Eq. ~(\ref{eq:Kappa}) by assuming a constant phonon lifetime, $\tau(\omega)=12$ ps, 
estimated from the upper limit of the experimentally evidenced linewidth of TA phonons in the Ba-Ge-Ni clathrate. 
Moreover we replace $\omega_{max}$ by a cutoff frequency $\omega_{cut}$, which is chosen to be the frequency at which the LA phonon branch looses its acoustic character, \textit{i.e.} the frequency at which the PR 
of the acoustic mode strongly decreases (change from green to red in Figs.\,\ref{fig7} and \ref{fig7b}). 
The introduction of this cutoff frequency takes the effect of the rattling modes into accout, which are responsible for a further flattening of the optical branches, as was discussed in subsection \ref{ss:dispersion}.
Figure \,\ref{fig10} shows the frequency dependence of $\kappa_{LA}^{[011]}(\omega)$ along a LA branch at room temperature, as well as the resulting $\kappa(T)$ in the inset. The reduced thermal conductivity 
of Ge$_{46}$ with respect to pure Ge shows that the cage structure itself is significantly decreasing the lattice 
thermal conductivity. An additional reduction of $\kappa$ in Ba$_8$Ge$_{40}$Ni$_{6}$ arises from the low energy Ba rattling phonons which lower the cutoff energy. Note that if for pure Ge we use the same cutoff frequency 
as for Ge$_{46}$, 
this results, despite the differences in $v(\omega)$,  in an similarly low thermal conductivity, as can be seen from the dashed curve in Fig.\,\ref{fig10}. 
Thus, to lowest order, the thermal transport by acoustic phonons in the Ba-Ge-Ni-clathrates corresponds to that of a simple Debye system with the sound velocity of pure Ge and a cutoff frequency corresponding to the energy of 
the first rattling mode of the guest atoms.\\
It is of course true that to fully calculate the thermal properties of a material optical phonon modes also have to be taken into account, therefore our model calculation has to be understood as a first order
approximation. This is justified, since the main contribution to heat transport can be attributed to acoustic phonons, while the optical modes above the introduced cutoff frequency can be negelected in a first order approach 
due to smaller lifetimes and group velocities.

\section{III. CONCLUSION}

We have performed a high resolution INS study of the lattice dynamics of one of the prototypal thermoelectric clathrate systems, Ba-Ge-Ni. 
Our data exclude an interpretation in terms of an isolated oscillator, since coherent modes of guest and host system are evidenced. 
Furthermore 
our results point out that Umklapp scattering is not able to account for the reduced $\kappa_L$ in the Ba-Ge-Ni system. The evidenced phonon lifetimes are at 
least an order of magnitude bigger than 
what would be expected if such a low $\kappa_L$ was to be reached by Umklapp scattering processes. Therefore a new mechanism was suggested to explain the 
low lattice thermal conductivity in clathrates 
-- the phononic low-pass filtering of acoustic phonon modes, having its origin in both loosely bound guest atoms and structural complexity of the host framework.
We are confident that these new insights will promot further studies in clathrates and other guest-host materials.


\begin{acknowledgements}
We thank B. Hennion, retired from LLB, and S. Merabia, from the LPMCN, for discussions. This work was supported by the EU Network of Excellence on Complex Metallic Alloys (Grant No. NMP3-CT-2005-500140) and within the European C-MAC
network. L. N. and S. P. acknowledge support from the Austrian Science Fund (project P19458-N16).
\end{acknowledgements}

\end{document}